# On number-ratio fluctuations in high-energy particle-production

P. Christiansen, E. Haslum and E. Stenlund*
*Experimental High-Energy Physics, Lund University, Box 118, SE-221 00 Lund, Sweden*
*corresponding author: evert.stenlund@hep.lu.se

*Abstract*: In this paper we will discuss the previously proposed quantity $\nu_{dyn}$[1], as a measure of the number-ratio fluctuations in high-energy particle-production. We will show that $\nu_{dyn}$ has pleasing mathematical properties making it ideal for the purpose. We will demonstrate its relation to two-particle correlations and how this measure can be generalized to higher order correlations.

1. Introduction

At high enough energies nuclear matter is supposed to transform into a state where nucleons are no longer the basic constituents, but rather quarks and gluons. Phase transitions like this often show large fluctuations in various measurable quantities, like e.g. bubble sizes in boiling water. In high-energy interactions fluctuations can be studied in the variation in the number of produced particles of various species, or even better in ratios of such numbers. However, since the number of produced particles is limited, fluctuations are to a large extent of statistical nature and efforts have to be made in order to find out whether there are additional, dynamical fluctuations present or not. Pruneau, Gavin, and Voloshin [1] have shown that the fluctuation measure $\nu_{dyn}$ has robust properties, but it was not shown how one can learn about the underlying physics, i.e. how to interpret $\nu_{dyn}$.

In this paper we investigate how $\nu_{dyn}$ can be related to the underlying physics for two models; pair production and the case of two distinct event classes. We extend pair production to triplets and show that two- and three- particle correlations can be differentiated only if the $\nu_{dyn}$-value exceeds the maximum value for two-particle correlations. Some very important sum rules are shown to relate fluctuations when charge states are considered. Finally we discuss how the experimental analysis can affect $\nu_{dyn}$. In sections 2-4 we discuss some of the ideas already expressed in [1], but the sections 5-12 contain material never discussed before.

The paper is organized as follows: In section 2 we will shortly discus the width of the distribution of the event-by-event ratio as a measure of fluctuations. In section 3 the relative multiplicity difference will be introduced together with the fluctuation measure, $\nu_{dyn}$. In section 4 we will study the case where the particles are uncorrelated. In section 5 we will look at the case of two-particle correlations. Section 6 is discussing sum rules. In section 7 we will shortly look at higher order correlations. In section 8 we demonstrate that fluctuations can appear without introducing correlations. In section 9 we generalize the formalism to look for higher order effects. In section 10 we discuss the influence of a limited acceptance, and in section 11 we shortly discuss detector effects. Finally, in section 12, we give an expression for the statistical error of $\nu_{dyn}$.

2. Studies of the event-by-event ratio

The most straight forward way to study number-ratio fluctuations would be to study the event-by-event distribution of the ratio, $\frac{n}{m}$, where n is the number of particles of one species, and m is the number of particles of another species. The dispersion of this distribution, $\sigma_{meas}$, then serves as a measure of the fluctuations. However, since a dominant part of this dispersion originate from statistical sources, the statistical part of the dispersion, $\sigma_{stat}$, has to be estimated using event mixing techniques or other kind of methods. The dynamical part of the fluctuations can then be extracted as

$$\sigma_{dyn} = \text{sign}(\sigma_{meas} - \sigma_{stat}) \cdot \sqrt{|\sigma_{meas}^2 - \sigma_{stat}^2|} . \qquad (1)$$

Positive values of $\sigma_{dyn}$ indicate that the sample shows larger fluctuations than the purely statistical ones, whereas a negative value indicates that the statistical fluctuations are reduced by some mechanism.

This method has many drawbacks. Firstly one can show that inefficiencies, i.e. a random portion of the tracks in each event goes unobserved, change the value of $\sigma_{dyn}$ in a non trivial way. Secondly the ratio $\frac{m}{n}$ will not give the same results as the ratio $\frac{n}{m}$. Thirdly, events with m = 0 (the most extreme fluctuations) have to be discarded in the analysis. Furthermore, since dispersions are most sensitive to the tails of the distributions, the method has a tendency to be more sensitive to low-multiplicity events in the sample.

### 3. The variance of the relative multiplicity difference

A more promising way to study number-ratio fluctuations is to study the distribution of the relative multiplicity difference $\frac{m}{\langle m \rangle} - \frac{n}{\langle n \rangle}$ [1]. By definition the average value of this expression is zero. The corresponding variance of this quantity is

$$v = \left\langle \left( \frac{m}{\langle m \rangle} - \frac{n}{\langle n \rangle} \right)^2 \right\rangle = \frac{\langle m^2 \rangle}{\langle m \rangle^2} - 2\frac{\langle mn \rangle}{\langle m \rangle \langle n \rangle} + \frac{\langle n^2 \rangle}{\langle n \rangle^2}. \qquad (2)$$

As we will see below, for purely statistical fluctuations this expression is reduced to

$$v_{stat} = \frac{1}{\langle m \rangle} + \frac{1}{\langle n \rangle}, \qquad (3)$$

and thus we have

$$v_{dyn} = \frac{\langle m(m-1) \rangle}{\langle m \rangle^2} - 2\frac{\langle mn \rangle}{\langle m \rangle \langle n \rangle} + \frac{\langle n(n-1) \rangle}{\langle n \rangle^2}. \qquad (4)$$

Note that since $v \geq 0$, $v_{dyn} \geq - v_{stat}$. The first and third terms in eq. (4) we recognize as scaled factorial moments of the second order. For a Poisson distribution, with $\sigma^2 = \langle n \rangle$, we obtain $\frac{\langle n(n-1) \rangle}{\langle n \rangle^2} = 1$. For broader distributions we find $\frac{\langle n(n-1) \rangle}{\langle n \rangle^2} > 1$ and for narrower distributions we find $\frac{\langle n(n-1) \rangle}{\langle n \rangle^2} < 1$. Scaled factorial moments have the nice property of filtering out statistical noise [2]. This means that if we have $m_{true}$ particles in an event but only observe a certain fraction, $m_{obs}$, of them we have

$$\frac{\langle m_{obs}(m_{obs}-1)\rangle}{\langle m_{obs}\rangle^2} = \frac{\langle m_{true}(m_{true}-1)\rangle}{\langle m_{true}\rangle^2}. \quad (5)$$

This can be shown in the following way. Let $\pi_G(N)$ be a global multiplicity distribution. Let each particle be observed with a fixed probability, p, independent of the rest of the particles. For fixed multiplicity, N, the distribution of n, the number of observed particles, will be

$$\pi_N(n) = \binom{N}{n} p^n (1-p)^{N-n}$$

and the global distribution becomes

$$\pi_G(n) = \sum_N \pi_G(N) \binom{N}{n} p^n (1-p)^{N-n}$$

Now

$$\langle n \rangle = \sum_n n\, \pi_G(n) = \sum_N \pi_G(N) \sum_n n \binom{N}{n} p^n (1-p)^{N-n} =$$

$$= \sum_N \pi_G(N) \sum_n \frac{nN!}{n!(N-n)!} p^n (1-p)^{N-n} =$$

$$= \sum_N \pi_G(N) \sum_n pN \frac{(N-1)!}{(n-1)!(N-n)!} p^{n-1} (1-p)^{N-n} =$$

$$= \sum_N pN\, \pi_G(N) \sum_n \frac{(N-1)!}{(n-1)!(N-n)!} p^{n-1} (1-p)^{N-n} = \sum_N pN\, \pi_G(N) = p\langle N \rangle$$

and

$$\langle n(n-1)\rangle = \sum_N \pi_G(N) \sum_n n(n-1)\binom{N}{n} p^n (1-p)^{N-n} =$$

$$= \sum_N \pi_G(N) \sum_n \frac{n(n-1)N!}{n!(N-n)!} p^n (1-p)^{N-n} =$$

$$= \sum_N \pi_G(N) \sum_n p^2 N(N-1) \frac{(N-2)!}{(n-2)!(N-n)!} p^{n-2} (1-p)^{N-n}$$

$$= \sum_N p^2 N(N-1)\, \pi_G(N) = p^2 \langle N(N-1)\rangle$$

so that

$$\frac{\langle n(n-1)\rangle}{\langle n\rangle^2} = \frac{\langle N(N-1)\rangle}{\langle N\rangle^2}$$

In the same way a similar relation can be derived for the second term in eq. (4), namely

$$\frac{\langle m_{obs} n_{obs}\rangle}{\langle m_{obs}\rangle\langle n_{obs}\rangle} = \frac{\langle m_{true} n_{true}\rangle}{\langle m_{true}\rangle\langle n_{true}\rangle}. \quad (6)$$

Note that this relation holds true even if the two particle species are observed with different efficiencies.

## 4. The uncorrelated case

If we now consider the case when the two particle species are uncorrelated, i.e. each particle has a fixed probability, p, to be of species a, and consequently the probability, 1–p, to be of species b, and furthermore write m + n = M, we find in the same way as we derived eq. (5) that

$$\frac{\langle m(m-1) \rangle}{\langle m \rangle^2} = \frac{\langle M(M-1) \rangle}{\langle M \rangle^2} \quad \text{and} \quad \frac{\langle n(n-1) \rangle}{\langle n \rangle^2} = \frac{\langle M(M-1) \rangle}{\langle M \rangle^2} \quad (7)$$

It can also be shown that the second term in eq. (4) will have the same expectation value

$$\frac{\langle mn \rangle}{\langle m \rangle \langle n \rangle} = \frac{\langle m(M-m) \rangle}{\langle m \rangle \langle M-m \rangle} = \frac{\langle M(M-1) \rangle}{\langle M \rangle^2} \quad (8)$$

Let $\pi_G(M)$ be the global multiplicity distribution. Now $\langle m \rangle = p\langle M \rangle$ and $\langle M-m \rangle = (1-p)\langle M \rangle$. Furthermore

$$\langle Mm \rangle = \sum_M M \, \pi_G(M) \sum_m m \binom{M}{m} p^m (1-p)^{M-m} = p\langle M^2 \rangle$$

and

$$\langle m(M-m) \rangle = \langle Mm \rangle - \langle m^2 \rangle = \langle Mm \rangle - \langle m(m-1) \rangle - \langle m \rangle =$$

$$= p\langle M^2 \rangle - p^2 \langle M(M-1) \rangle - p\langle M \rangle = (p-p^2)\langle M(M-1) \rangle = p(1-p)\langle M(M-1) \rangle.$$

Thus

$$\frac{\langle m(M-m) \rangle}{\langle m \rangle \langle M-m \rangle} = \frac{\langle M(M-1) \rangle}{\langle M \rangle^2}.$$

Combining these results we find that $\nu_{dyn} = 0$ for the case of uncorrelated particle production, independent of the underlying multiplicity distribution. We also find that the statistical part of eq. (2) is given by eq. (3). Furthermore we see that $\nu_{dyn}(m,n) = \nu_{dyn}(n,m)$, due to the symmetric form of eq. (4).

From the discussion so far it should be clear that $\nu_{dyn}$ is independent of efficiencies and on the underlying multiplicity distribution, and thus doesn't suffer from any of the deficiencies associated with $\sigma_{dyn}$ above.

## 5. The effect of two-particle correlations

Non-statistical fluctuations can arise in a sample in different ways. Let's first look at the effects introduced by two-particle correlations.

Let's assume that a random particle is of species a with probability p and of species b with probability q = 1 – p. Let us furthermore assume that all particles are produced in pairs. There will then be three different kind of pairs; aa, ab and bb. In the uncorrelated case the corresponding probabilities will be $p^2$, 2pq, and $q^2$, respectively. Two-particle correlations can now be introduced by changing these probabilities to $p^2 + \varepsilon$, $2pq - 2\varepsilon$, and $p^2 + \varepsilon$, respectively. Note that this is the only parameterization that ensures that the one particle probabilities come out correctly. We can now calculate $v_{dyn}(a,b)$ in this case letting m denote the multiplicity of species a, and n denote the multiplicity of species b. Furthermore we have $m + n = M$. When we now consider a random pair of particles in an event with multiplicity M, we will have $\frac{M(M-1)}{2}$ such possible pairs. Out of these pairs $\frac{M}{2}$ come from truly correlated pairs, and $\frac{M(M-2)}{2}$ come from random uncorrelated pairs. We find

$$\langle m(m-1) \rangle = \langle M(p^2 + \varepsilon) + M(M-2)p^2 \rangle = \langle M(M-1) \rangle p^2 + \langle M \rangle \varepsilon,$$

$$\langle n(n-1) \rangle = \langle M(q^2 + \varepsilon) + M(M-2)q^2 \rangle = \langle M(M-1) \rangle q^2 + \langle M \rangle \varepsilon, \text{ and}$$

$$\langle mn \rangle = \langle M(pq - \varepsilon) + M(M-2)pq \rangle = \langle M(M-1) \rangle pq - \langle M \rangle \varepsilon.$$

These results combine to

$$v_{dyn} = \frac{\langle m(m-1) \rangle}{\langle m \rangle^2} - 2\frac{\langle mn \rangle}{\langle m \rangle \langle n \rangle} + \frac{\langle n(n-1) \rangle}{\langle n \rangle^2} =$$

$$= \frac{\langle M(M-1) \rangle}{\langle M \rangle^2} + \frac{\varepsilon}{\langle M \rangle p^2} - 2\frac{\langle M(M-1) \rangle}{\langle M \rangle^2} + 2\frac{\varepsilon}{\langle M \rangle pq} + \frac{\langle M(M-1) \rangle}{\langle M \rangle^2} + \frac{\varepsilon}{\langle M \rangle q^2},$$

i.e.

$$v_{dyn} = \frac{\varepsilon}{\langle M \rangle p^2 q^2}. \tag{9}$$

This result shows that $v_{dyn}$ is a direct measurement of the two-particle correlation parameter, $\varepsilon$. Since $\varepsilon \leq pq$ we have $v_{dyn} \leq \frac{1}{\langle M \rangle pq}$ if two-particle correlations is the only source of fluctuations. Also $\varepsilon \geq -\min(p^2,q^2)$ gives $v_{dyn} \geq -\frac{1}{\langle M \rangle \max(p^2,q^2)}$. With $p = q = \frac{1}{2}$ this relation becomes $v_{dyn} \geq -\frac{4}{\langle M \rangle}$, a result also obtained in [1]. This case corresponds to damped fluctuations induced by various conservation laws, e.g. electrical charge, strangeness, baryon number, *etc.*, and in such cases dynamical fluctuations of this kind will dominate. Note however, whereas $v_{dyn}$ automatically is corrected for inefficiencies, the right side of eq. (9) is not. Thus $\langle M \rangle$, p, and q have to be corrected in order to get an efficiency corrected value of $\varepsilon$.

Eq. (9) also indicates that if two-particle correlations are the same, independent of multiplicity, $v_{dyn}$ should decrease as a function of multiplicity due to the $\frac{1}{\langle M \rangle}$-factor. Regarding K-to-π fluctuations, an approximate $\frac{1}{\langle M \rangle}$-dependence has been observed in recent Au+Au data at 200 and 62.4 GeV from the STAR-collaboration [3].

6. **Sum rules**

If one of the species has two charge states, $a_+$ and $a_-$, one can apart from $v_{dyn}(a,b)$ also calculate $v_{dyn}(a_+,b)$ and $v_{dyn}(a_-,b)$ as well as $v_{dyn}(a_+,a_-)$. There is however a relationship between those quantities. Let $m_+$ and $m_-$ be the respective number of particles of the two charge states of species a, and n be the number of particles of species b. We have $m_+ + m_- = m$ and $\tilde{\alpha} = \frac{\langle m_+ \rangle}{\langle m \rangle}$.

Now

$$v_{dyn}(a,b) = \frac{\langle (m_++m_-)(m_++m_--1) \rangle}{\langle m \rangle^2} - 2\frac{\langle (m_++m_-)n \rangle}{\langle m \rangle \langle n \rangle} + \frac{\langle n(n-1) \rangle}{\langle n \rangle^2} =$$

$$= \tilde{\alpha}^2 \frac{\langle m_+(m_+-1) \rangle}{\langle m_+ \rangle^2} + (1-\tilde{\alpha})^2 \frac{\langle m_-(m_--1) \rangle}{\langle m_- \rangle^2} + 2\tilde{\alpha}(1-\tilde{\alpha})\frac{\langle m_+m_- \rangle}{\langle m_+ \rangle \langle m_- \rangle}$$

$$-2\tilde{\alpha}\frac{\langle m_+n \rangle}{\langle m_+ \rangle \langle n \rangle} - 2(1-\tilde{\alpha})\frac{\langle m_-n \rangle}{\langle m_- \rangle \langle n \rangle} + \frac{\langle n(n-1) \rangle}{\langle n \rangle^2},$$

which reduces to

$$v_{dyn}(a,b) = \tilde{\alpha} v_{dyn}(a_+,b) + (1-\tilde{\alpha})v_{dyn}(a_-,b) - \tilde{\alpha}(1-\tilde{\alpha})v_{dyn}(a_+,a_-). \qquad (10)$$

This result shows that one of the contributions to positive (enhanced) fluctuations in the a-to-b ratio comes from negative (reduced) fluctuations in the $a_+$-to-$a_-$ ratio. It should be noted that quantities like $v_{dyn}(a_+,a_-)$ generally are negative, due to charge conservation and neutral decays.

If also species b has two charge states we obtain

$$v_{dyn}(a,b) = \tilde{\alpha}\tilde{\beta} v_{dyn}(a_+,b_+) + (1-\tilde{\alpha})\tilde{\beta} v_{dyn}(a_-,b_+) + \tilde{\alpha}(1-\tilde{\beta})v_{dyn}(a_+,b_-)$$

$$+ (1-\tilde{\alpha})(1-\tilde{\beta})v_{dyn}(a_-,b_-) - \tilde{\alpha}(1-\tilde{\alpha})v_{dyn}(a_+,a_-) - \tilde{\beta}(1-\tilde{\beta})v_{dyn}(b_+,b_-), \qquad (11)$$

where $\tilde{\beta} = \frac{\langle n_+ \rangle}{\langle n \rangle}$ with $n = n_+ + n_-$. If $\tilde{\alpha} \approx \tilde{\beta} \approx \frac{1}{2}$ (which is the common case) then

$$v_{dyn}(a,b) \approx \frac{1}{4}\Big[v_{dyn}(a_+,b_+) + v_{dyn}(a_-,b_+) + v_{dyn}(a_+,b_-) + v_{dyn}(a_-,b_-)$$

$$- v_{dyn}(a_+,a_-) - v_{dyn}(b_+,b_-)\Big]. \qquad (12)$$

This result is in accordance with recent studies by the STAR-collaboration [3] who found

$$v_{dyn}(K,\pi) > v_{dyn}(K_+,\pi_+) \approx v_{dyn}(K_-,\pi_-) > v_{dyn}(K_+,\pi_-) \approx v_{dyn}(K_-,\pi_+)$$

in Au+Au data at RHIC-energies.

## 7. Higher order correlations

One can also consider three-particle correlations. We now have three possible triplet combinations; aaa, aab, abb, and bbb. In the uncorrelated case the corresponding probabilities will be $p^3$, $3p^2q$, $3pq^2$ and $q^3$. Three-particle correlations are introduced by changing these probabilities to $p^3 + \alpha$, $3p^2q - 2\alpha + \beta$, $3pq^2 + \alpha - 2\beta$ and $q^3 + \beta$, respectively, again conserving the one-particle probabilities. After some calculations it is found that

$$\nu_{dyn} = \frac{2(\alpha + \beta)}{3\langle M \rangle p^2 q^2}. \tag{13}$$

When we compare this expression with eq. (9) we find that, with $\alpha + \beta = \frac{3}{2}\varepsilon$ the two expressions are the identical. If, however, three-particle correlations are at work $\nu_{dyn}$-values exceeding the two-particle correlation limit, obtained above (after eq. (9)), can be reached. E.g. with $\alpha = p - p^3$, and $\beta = q - q^3$, i.e. the case when only aaa and bbb triplets are produced, we obtain $\alpha + \beta = 3pq$ and

$$\nu_{dyn} = \frac{2}{\langle M \rangle pq},$$

i.e. twice the limit for two-particle correlations. This we expect since each particle now belongs to two different correlated pairs. In the general case of n-particle correlations, the limit will be proportional to $(n - 1)$. This means that very large fluctuations, $\nu_{dyn} > \frac{1}{\langle M \rangle pq}$, is a clear indication of multi-particle correlations.

### 8. Fluctuations without correlations

Another possible source of fluctuations is the mixing of two slightly different samples, each of them without fluctuations. Let's assume that we have two uncorrelated samples with the same multiplicity distribution. Sample 1 has the overall fraction $\xi$ of the events, and sample 2 has the overall fraction $1 - \xi$. Furthermore the first sample has the probability $p_1$ for particle species a, and $q_1 = 1 - p_1$ for species b, and the second sample has the probability $p_2$ for particle species a, and $q_2 = 1 - p_2$ for species b. In this case each sample shows no fluctuations, but the mixing introduces fluctuations in the final sample. When $\nu_{dyn}$ is calculated for this case we find

$$\nu_{dyn} = \frac{\langle M(M-1) \rangle}{\langle M \rangle^2} \frac{\xi(1-\xi)(\Delta p)^2}{\bar{p}^2 \bar{q}^2}, \tag{14}$$

where M is the total multiplicity (species a + b), $\bar{p} = \xi p_1 + (1-\xi)p_2$, $\bar{q} = \xi q_1 + (1-\xi)q_2$, and $\Delta p = |p_1 - p_2|$. Note that the multiplicity dependence in eq. (14) is very different from the one obtained in eq. (9), and that $\nu_{dyn}$ is non-negative. In the case where the two multiplicity distributions are not the same, $\nu_{dyn}$ will be non-zero only in the overlap regions (see fig. 1). The underlying physics behind this idea could be realized, if two phases are present in the final state of the system under study. Note that this kind of effect should be particularly easy to observe at large multiplicities.

### 9. Generalization of $\nu_{dyn}$

It is possible to generalize eq. (2) to measure higher order fluctuations. The expression for third order dynamical fluctuations will be

$$\nu_3 = \frac{\langle m(m-1)(m-2)\rangle}{\langle m\rangle^3} - 3\frac{\langle m(m-1)n\rangle}{\langle m\rangle^2\langle n\rangle} + 3\frac{\langle mn(n-1)\rangle}{\langle m\rangle\langle n\rangle^2} - \frac{\langle n(n-1)(n-2)\rangle}{\langle n\rangle^3}. \qquad (15)$$

From this expression we se that $\nu_3(a,b) = -\nu_3(b,a)$. Furthermore we find that for two-particle correlations, parameterized as before, $\nu_3 = 0$, independent of $\varepsilon$. For three-particle correlations, with the parameters $\alpha$ and $\beta$ from above, we find

$$\nu_3 = \frac{2(q\alpha - p\beta)}{\langle M\rangle^2 p^3 q^3}$$

and with $\alpha = \frac{3}{2}p\varepsilon + \delta$ and $\beta = \frac{3}{2}q\varepsilon - \delta$, ensuring $\alpha + \beta = \frac{3}{2}\varepsilon$, this expression reduces to

$$\nu_3 = \frac{2\delta}{\langle M\rangle^2 p^3 q^3}, \qquad (16)$$

where $\delta$ can be called the asymmetry parameter. Note that whenever the asymmetry parameter is zero, two- and three-particle correlations cannot be distinguished from each other, unless $\nu_{dyn}$ is outside the limits for two-particle correlations. Furthermore three-particle correlation effects in eq. (16) behave as $\frac{1}{\langle M\rangle^2}$, whereas two-particle correlation effects in eq. (9) behave as $\frac{1}{\langle M\rangle}$. This indicates that, in high-energy heavy-ion experiments, where the multiplicities can be very large, three-particle correlation effects might be hard to observe.

### 10. The influence of limited acceptance

Up to now we have only discussed inefficiencies in the meaning fixed-probability random losses of particles. A related question is how $\nu_{dyn}$ is influenced by a decrease in acceptance. In fig. 2 we have schematically shown three different possibilities of two-particle azimuthal correlations. In fig. 2a the particle pairs are produced close in azimuth and a smaller azimuthal acceptance will essentially either keep both particles from a pair inside the acceptance or both particles will be lost. In this case $\varepsilon$ will be the same inside the acceptance window as for full acceptance, and $\nu_{dyn}$ will be inversely proportional to the multiplicity inside the window.

In fig. 2b the particle pairs are produced back-to-back and a decrease of acceptance will essentially break all pairs and $\nu_{dyn} \rightarrow 0$. In fig. 2c particles within a pair are uncorrelated in $\varphi$, and a decrease of acceptance will have the same effect as a decrease of efficiency, i.e. $\nu_{dyn}$ will be unaffected by the change in acceptance.

Obviously, by changing the experimental acceptance, further information about the nature of the fluctuations can be gained.

### 11. Detector effects

As $\nu_{dyn}$ is very sensitive to small fluctuations, it also has a tendency to pick up effects originating from the detector itself. Such effects can favourably be studied by the Monte-Carlo technique. As an example, fig. 3 shows one such study. Here K-to-$\pi$ fluctuations are generated in events with 100 particles; 10% kaons and 90% pions with the two-particle correlation parameter ε = 8.1%. In fig. 3a a certain fraction of the kaons are misidentified as pions and in fig. 3b a certain fraction of the pions are misidentified as kaons. As can be seen, in this case it is more important to make sure that there are no misidentified pions among the kaons than *vice versa*. If 1% of the pions are misidentified as kaons the signal drops by around 15%. If 10% of the kaons are misidentified as pions the signal drops by around 2%. Another effect of importance is the risk of double-counting particles in an experiment. In fig. 4a a certain fraction of the kaons are double-counted and in fig. 4b the same thing is shown for pions. If 1% of the kaons are double-counted the signal increases by around 2%, whereas if 10% of the pions are double-counted, the increase is about the same. Again, in this asymmetric case, the signal is affected much more by effects that influence the kaons (the rarer species).

## 12. Statistical error of $\nu_{dyn}$

Even in present experiments the amount of data is limited. An estimation of $\nu_{dyn}$ has an experimental uncertainty. This uncertainty can be estimated using different methods. The data sample can be divided into several sub samples, and the uncertainty can be estimated from the variation in the results from the different sub samples. A Monte-Carlo model of the fluctuations can be constructed (e.g. from two-particle correlations with a suitable value of ε) and the variation in the results from several runs can be calculated. A third method is to use an analytic expression. Thus we finally give our results to the leading order for the variance in $\nu_{dyn}$, where m and n are the numbers of the two species, respectively, and N is the number of events in the sample.

$$V[\nu_{dyn}] = \frac{1}{N} \frac{1}{\langle m\rangle^6 \langle n\rangle^6} \{ 6\langle n^2 m^2\rangle\langle n\rangle^4\langle m\rangle^4 - 4\langle n^3 m\rangle\langle n\rangle^3\langle m\rangle^5 - 4\langle nm^3\rangle\langle n\rangle^5\langle m\rangle^3$$

$$+ 8\langle n^2 m\rangle\langle n^2\rangle\langle n\rangle^2\langle m\rangle^5 + 8\langle nm^2\rangle\langle m^2\rangle\langle n\rangle^5\langle m\rangle^2 - 4\langle n^2 m\rangle\langle nm\rangle\langle n\rangle^3\langle m\rangle^4$$

$$- 4\langle nm^2\rangle\langle nm\rangle\langle n\rangle^4\langle m\rangle^3 - 4\langle n^2 m\rangle\langle m^2\rangle\langle n\rangle^4\langle m\rangle^3 - 4\langle nm^2\rangle\langle n^2\rangle\langle n\rangle^3\langle m\rangle^4$$

$$+ 8\langle nm\rangle^3\langle n\rangle^3\langle m\rangle^3 - 4\langle nm\rangle^2\langle n^2\rangle\langle n\rangle^2\langle m\rangle^4 - 4\langle nm\rangle^2\langle m^2\rangle\langle n\rangle^4\langle m\rangle^2 - 4\langle nm\rangle^2\langle n\rangle^4\langle m\rangle^4$$

$$+ 4\langle nm\rangle\langle n^3\rangle\langle n\rangle^2\langle m\rangle^5 + 4\langle nm\rangle\langle m^3\rangle\langle n\rangle^5\langle m\rangle^2 - 8\langle nm\rangle\langle n^2\rangle^2\langle n\rangle\langle m\rangle^5$$

$$- 8\langle nm\rangle\langle m^2\rangle^2\langle n\rangle^5\langle m\rangle + 8\langle nm\rangle\langle n^2\rangle\langle m^2\rangle\langle n\rangle^3\langle m\rangle^3 + 4\langle nm\rangle\langle n^2\rangle\langle n\rangle^3\langle m\rangle^5$$

$$+ 4\langle nm\rangle\langle m^2\rangle\langle n\rangle^5\langle m\rangle^3 + \langle n^4\rangle\langle n\rangle^2\langle m\rangle^6 + \langle m^4\rangle\langle n\rangle^6\langle m\rangle^2 - 4\langle n^3\rangle\langle n^2\rangle\langle n\rangle\langle m\rangle^6$$

$$- 4\langle m^3\rangle\langle m^2\rangle\langle n\rangle^6\langle m\rangle + 4\langle n^2\rangle^3\langle m\rangle^6 + 4\langle m^2\rangle^3\langle n\rangle^6 - \langle n^2\rangle^2\langle n\rangle^2\langle m\rangle^6 - \langle m^2\rangle^2\langle n\rangle^6\langle m\rangle^2$$

$$- 2\langle n^2\rangle\langle m^2\rangle\langle n\rangle^4\langle m\rangle^4 + 2\langle n^2 m\rangle\langle n\rangle^4\langle m\rangle^4 + 2\langle nm^2\rangle\langle n\rangle^4\langle m\rangle^4 - 4\langle n^2 m\rangle\langle n\rangle^3\langle m\rangle^5$$

$$- 4\langle nm^2\rangle\langle n\rangle^5\langle m\rangle^3 + 4\langle nm\rangle^2\langle n\rangle^4\langle m\rangle^3 + 4\langle nm\rangle^2\langle n\rangle^3\langle m\rangle^4 - 4\langle nm\rangle\langle n^2\rangle\langle n\rangle^3\langle m\rangle^4$$

$$- 4\langle nm\rangle\langle m^2\rangle\langle n\rangle^4\langle m\rangle^3 + 4\langle nm\rangle\langle n^2\rangle\langle n\rangle^2\langle m\rangle^5 + 4\langle nm\rangle\langle m^2\rangle\langle n\rangle^5\langle m\rangle^2$$

$$- 4\langle nm\rangle\langle n\rangle^5\langle m\rangle^4 - 4\langle nm\rangle\langle n\rangle^4\langle m\rangle^5 + 2\langle n^3\rangle\langle n\rangle^2\langle m\rangle^6 + 2\langle m^3\rangle\langle n\rangle^6\langle m\rangle^2$$

$$- 4\langle n^2\rangle^2\langle n\rangle\langle m\rangle^6 - 4\langle m^2\rangle^2\langle n\rangle^6\langle m\rangle + 2\langle n^2\rangle\langle n\rangle^4\langle m\rangle^5 + 2\langle m^2\rangle\langle n\rangle^5\langle m\rangle^4$$

$$+ 2\langle n^2\rangle\langle n\rangle^3\langle m\rangle^6 + 2\langle m^2\rangle\langle n\rangle^6\langle m\rangle^3 + 2\langle nm\rangle\langle n\rangle^4\langle m\rangle^4 + \langle n^2\rangle\langle n\rangle^2\langle m\rangle^6$$

$$+ \langle m^2\rangle\langle n\rangle^6\langle m\rangle^2 - \langle n\rangle^6\langle m\rangle^4 - \langle n\rangle^4\langle m\rangle^6 - 2\langle n\rangle^5\langle m\rangle^5 \} + \mathcal{O}(\frac{1}{N^2})$$

## 13. Summary

In this paper we have shown that the fluctuation measure $\nu_{dyn}$ is ideal for studies of number-ratio fluctuations in high-energy particle-production. We have shown that purely statistical fluctuations lead to $\nu_{dyn} = 0$, and that $\nu_{dyn}$ directly measures the two-particle parameter, $\varepsilon$. We have furthermore shown that higher-order correlations will be undistinguishable from two-particle correlations unless they are large enough to pass the limit imposed by two-particle correlations. A generalization of $\nu_{dyn}$ is shown to be sensitive to the asymmetric part of three-particle correlations. A relation between different fluctuations (sum rules) has been derived and an expression for the statistical error of $\nu_{dyn}$ is given. Detector effects as well as of acceptance limitations have been briefly discussed.

We have demonstrated that in searches for dynamical fluctuations it is advantageous to go to large multiplicities, M, where two-particle effects (including charge conservation) are suppressed by $1/\langle M\rangle$, whereas fluctuations originating from a mixture of different phases aren't suppressed at all.

**Acknowledgement**

The authors wish to express their gratitude to the Swedish Research Council for financial support.

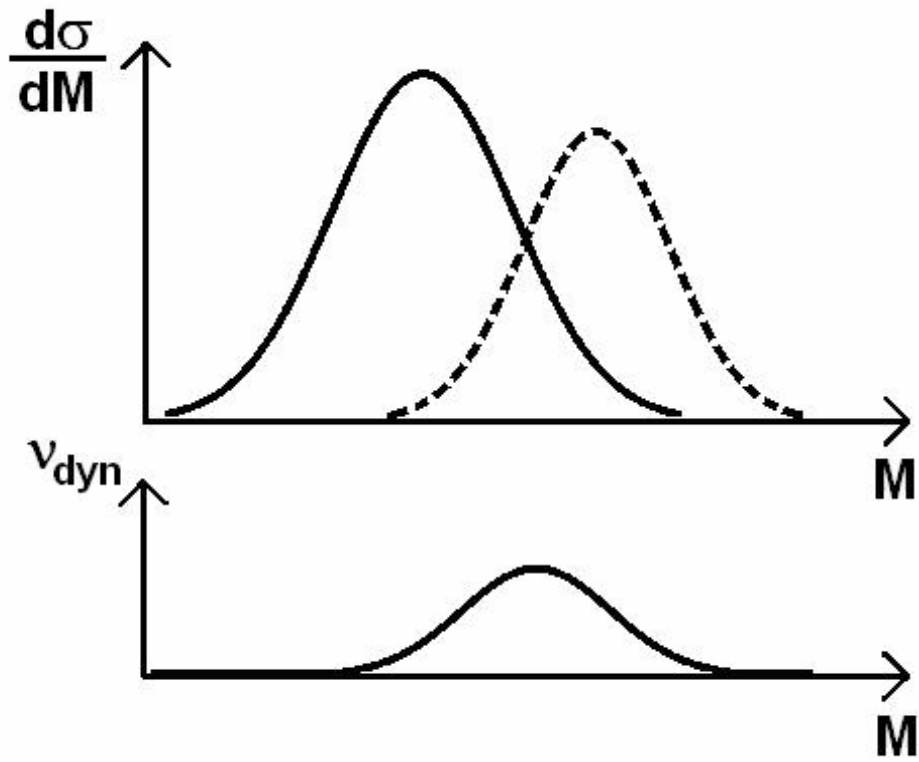

Figure 1. In the case where two uncorrelated samples are mixed, $\nu_{dyn}$ will be non-zero only in the regions where the multiplicity overlaps.

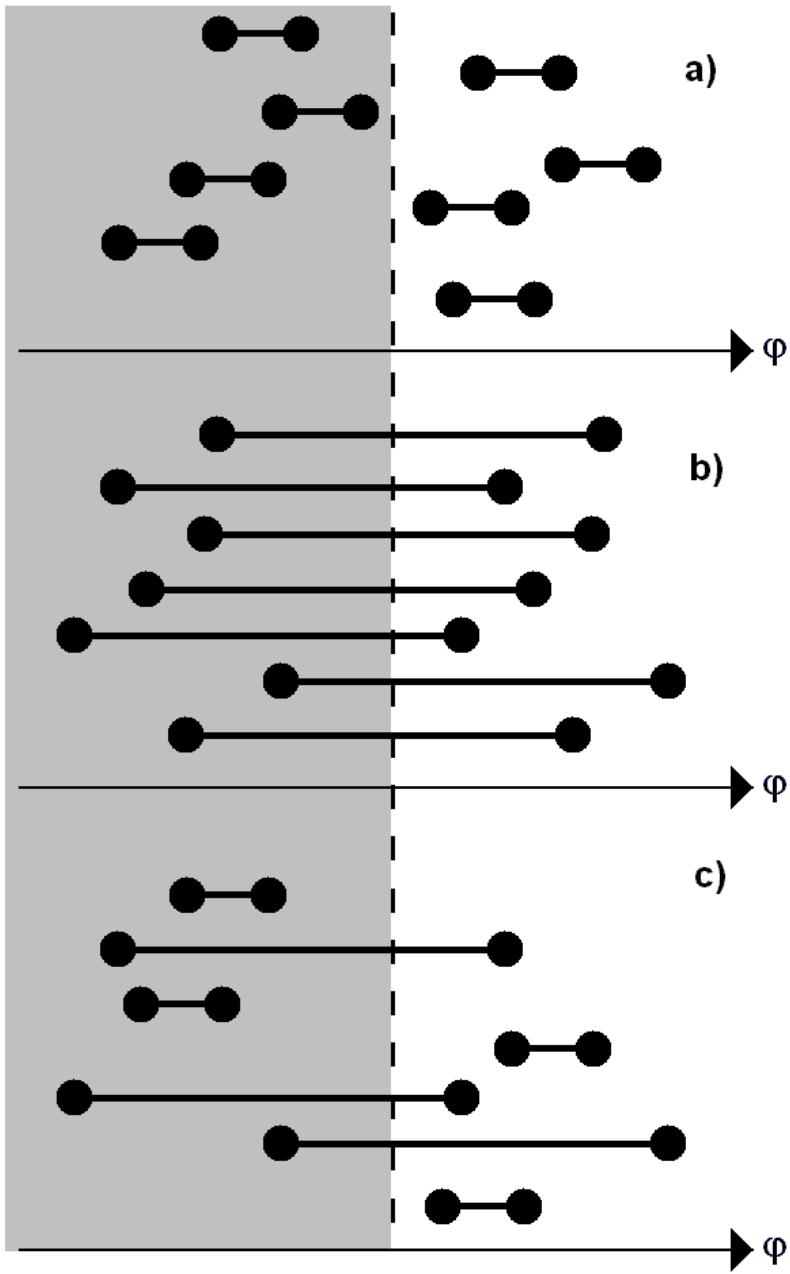

Figure 2. Various possibilities of two-particle azimuthal correlations.

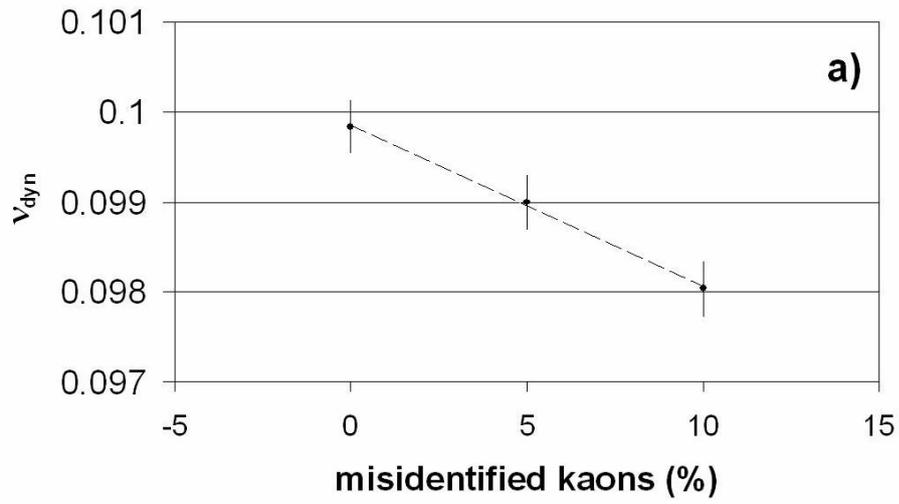

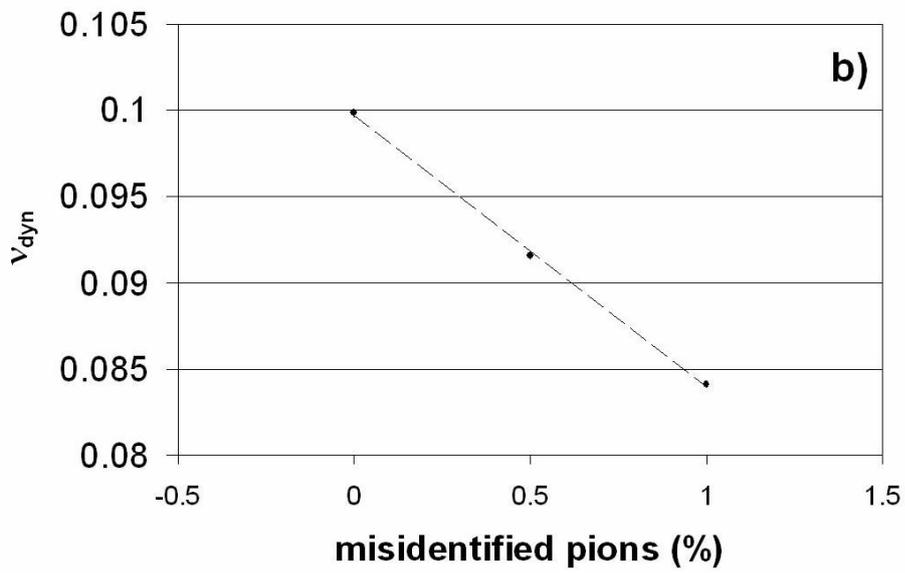

Figure 3. Simulation of the influence of particle misidentification.

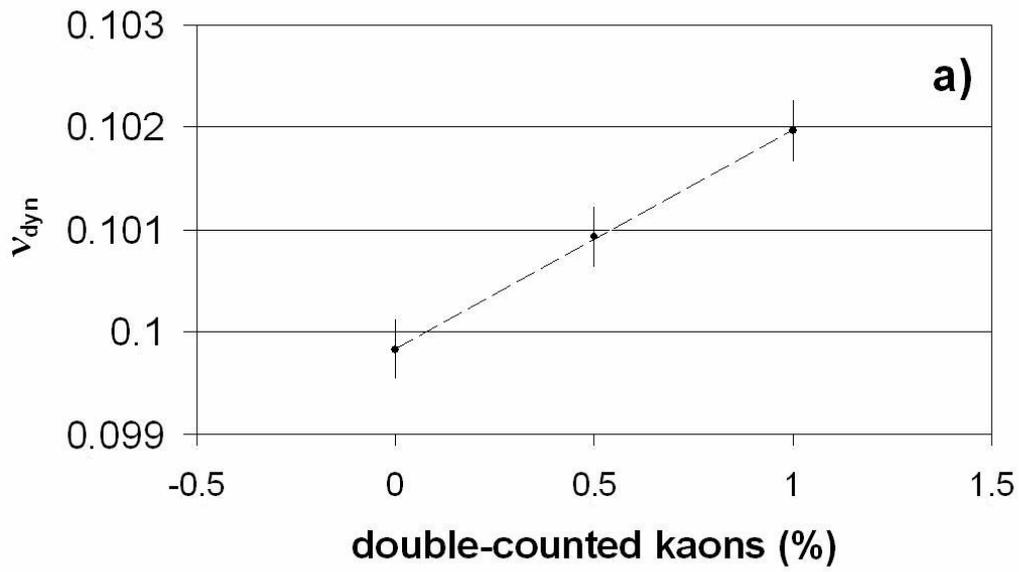

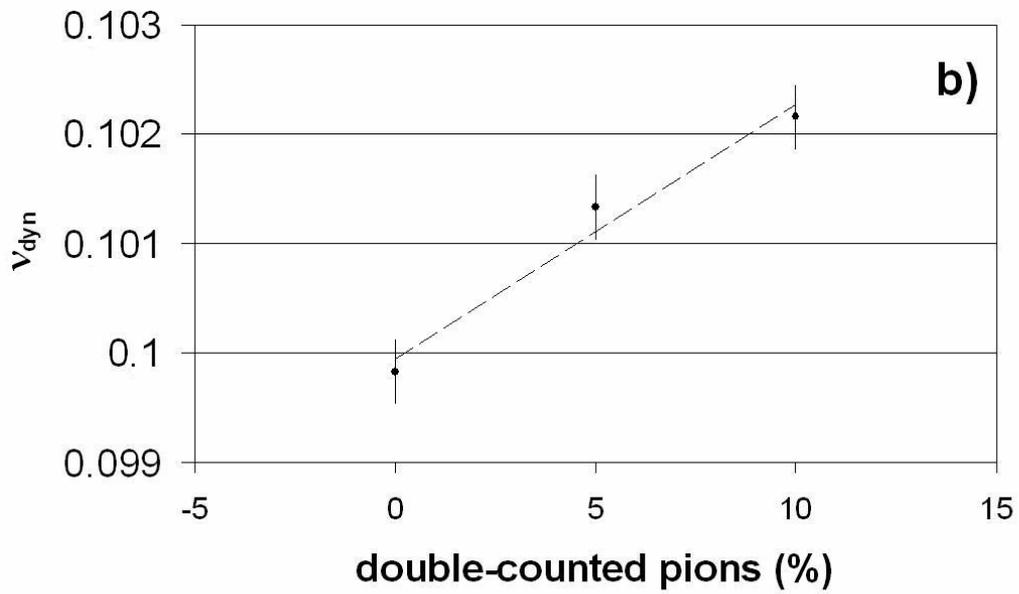

Figure 4. Simulation of the influence of double-counting.